\documentclass[pdflatex,sn-mathphys-num]{sn-jnl}

\usepackage{graphicx}%
\usepackage{multirow}%
\usepackage{amsmath,amssymb,amsfonts}%
\usepackage{amsthm}%
\usepackage{mathrsfs}%
\usepackage[title]{appendix}%
\usepackage{xcolor}%
\usepackage{textcomp}%
\usepackage{manyfoot}%
\usepackage{booktabs}%
\usepackage{listings}
\usepackage{tikz}
\usetikzlibrary{shapes.geometric, arrows.meta, positioning, calc}
\usepackage{array}%
\usepackage[capitalise]{cleveref}
\Crefname{equation}{Eq.}{Eqs.}
\Crefname{figure}{Fig.}{Figs.}
\Crefname{table}{Tab.}{Tabs.}
\Crefname{section}{Sec.}{Secs.}
\newcommand{\ket}[1]{\left|#1\right\rangle}

\begin{document}

\title[Quantum Feature Space of a Qubit Coupled to a Classical Stochastic Environment]{Quantum Feature Space of a Qubit Coupled to a Classical Stochastic Environment}

\author*[1]{\fnm{Chris} \sur{Wise}}\email{c.wise@unswalumni.com}
\author[2]{\fnm{Akram} \sur{Youssry}}
\author[2,3]{\fnm{Alberto} \sur{Peruzzo}}
\author[1]{\fnm{Jo} \sur{Plested}}
\author[4]{\fnm{Matt} \sur{Woolley}}
\affil[1]{\orgdiv{School of Systems and Computing}, \orgname{UNSW}, \orgaddress{\city{Canberra}, \state{ACT}, \postcode{2601}, \country{Australia}}}
\affil[2]{\orgdiv{Quantum Photonics Laboratory and Centre for Quantum Computation and Communication Technology}, \orgname{RMIT University}, \orgaddress{\city{Melbourne}, \state{VIC}, \postcode{3000}, \country{Australia}}}
\affil[3]{\orgname{Quandela}, \orgaddress{\city{Massy}, \country{France}}}
\affil[4]{\orgdiv{School of Engineering and Technology and Centre for Engineered Quantum Systems}, \orgname{UNSW}, \orgaddress{\city{Canberra}, \state{ACT}, \postcode{2601}, \country{Australia}}}
\abstract{
Qubit control protocols traditionally characterise qubit-environment coupling using resource-intensive pulse sweeps on quantum hardware. Although previous work attempted to reduce this quantum runtime by inferring noise operators, it relied on a complex grey-box approach combining deep neural networks with physics-encoded layers, posing challenges in training and scaling. Here, we introduce a \textit{quantum feature space} (QFS, denoted $\mathcal{F}$) constructed from a single pulse sequence. By engineering $\mathcal{F}$ to encode qubit-environment dynamics, we eliminate the need for expensive neural networks to learn these physical interactions. This allows us to offload extraction of spectral features to efficient classical machine learning algorithms. Bypassing both the extensive pulse sweeps of traditional quantum noise spectroscopy and the costly neural networks of prior methods, we demonstrate that a simple random forest algorithm on $\mathcal{F}$ classifies broad noise classes with $\sim 90\%$ accuracy under simulated realistic hardware constraints, including pulse jitter and readout errors. Furthermore, Euclidean distance searches over $\mathcal{F}$ can determine spectral resonances. Focusing on noise classification and identification rather than continuous power spectral density reconstruction makes $\mathcal{F}$ an efficient diagnostic tool for quantum hardware.
}


\maketitle

\section{Introduction}
Noise in quantum computers that results in qubits losing coherence (decoherence) is due to unwanted environment coupling~\cite{bravyi2018correcting}. One way to mitigate decoherence is to decouple the qubit from the environment by applying a control field in the form of a series of pulses~\cite{viola1999dynamical}. This technique originated in nuclear magnetic resonance~\cite{carr1954effects, meiboom1958modified, degen2017quantum, boss2016one}. Within this framework, $N_p$ refocusing $\pi$-pulses are applied separated by free-evolution delays, forming the Carr--Purcell--Meiboom--Gill (CPMG) sequence~\cite{carr1954effects, meiboom1958modified}.

\'Alvarez and Suter (AS) established dynamical decoupling noise spectroscopy by applying CPMG sequences with varied inter-pulse delays to sample the environmental spectrum~\cite{alvarez2011measuring}. In traditional spectroscopy protocols, empirical coherence data obtained from dynamical decoupling sequences are either inverted via filter-function matrices~\cite{alvarez2011measuring, von2020two, sun2022self} or fitted to theoretical coherence curves corresponding to parametric noise spectra. However, this method requires an accurate model of qubit dynamics~\cite{wise2021using}, which can be challenging to develop in general~\cite{martina2023machine}. Furthermore, if an inappropriate parametric noise spectrum is chosen \textit{a priori}, the estimated noise spectrum will correspond to an inaccurate model of qubit dynamics, and subsequently, control pulses optimised using this qubit model will likely be suboptimal~\cite{martina2023machine}.

More direct methods of noise spectroscopy include techniques such as two-point correlation, which measures the correlation function of the noise by implementing gates at different times and measuring the change in the qubit state~\cite{szankowski2017, von2020two}. The Fourier transform of this correlation function yields the noise spectrum. However, this method can be resource-intensive, time-consuming, computationally expensive, and requires that the noise be stationary~\cite{von2020two,wise2021using,martina2023deep,vezvaee2024fourier}.

Vezvaee et al.~\cite{vezvaee2024fourier} developed Fourier transform noise spectroscopy (FTNS), which applies the Fourier transform to free-induction decay (FID) coherence measurements. The technique accurately recovered the correct noise spectra and outperformed previous decoupling schemes while significantly reducing experimental overhead. However, the method makes several assumptions that limit its applicability. For example, the authors assumed that the qubit is only subject to dephasing, where the qubit thermal relaxation process occurs over a much longer time scale than phase randomisation, and that the frequency fluctuations of a qubit are subject to stationary zero-mean Gaussian noise.

Instead of analytical methods, Martina et al.~\cite{martina2023deep} demonstrated the effectiveness of deep learning by using it to reconstruct the power spectral density (PSD) of an ensemble of carbon impurities around a diamond's nitrogen-vacancy centre.

Wise et al.~\cite{wise2021using} also used deep learning techniques to infer accurate qubit noise spectra. The method used measurement data obtained from CPMG coherence decay curves as input to a recurrent neural network with a feedforward neural network head to estimate the noise spectrum. They showed that the model could accurately estimate the qubit noise spectrum using only coherence curves and outperformed the standard noise spectroscopy techniques.

Beyond these model-specific limitations, an overarching bottleneck of traditional and deep learning approaches is that constructing full coherence decay curves still requires sweeping evolution times or inter-pulse delays across many separate sequences at inference time. This exhaustive sampling can be expensive in terms of quantum computer resources, posing a challenge for the frequent calibration required by noisy intermediate-scale quantum (NISQ) devices, as noise processes in qubits can drift over time~\cite{seedhouse2025wavelet,nakamura2024gate}. To circumvent the need for full coherence curve sweeps, researchers have explored alternative architectures that operate on individual control sequences.

Youssry et al.~\cite{youssry2020characterization} introduced a grey-box model to separate control and qubit-environment dynamics. In this work, time-dependent Hamiltonians and unitaries specified the control dynamics, and a recurrent neural network was used to predict the qubit-environment dynamics associated with control pulses. Once trained on a library of pulse sequences, such a model requires only a single arbitrary control sequence as input to predict the corresponding noisy evolution.

Despite the success of such frameworks~\cite{youssry2020characterization,youssry2022multi}, these approaches are ultimately limited by their use of deep learning models. These models are often large and require extensive datasets, making them expensive to train. This is undesirable due to the drift in noise dynamics, which necessitates frequent retraining. Additionally, noise profiles may change from one physical qubit to another~\cite{sung2019non}, requiring qubit-specific models, which do not scale efficiently. Furthermore, specific to the grey-box method of~\cite{youssry2020characterization,youssry2022multi}, the white-box aspect scales exponentially with the number of qubits.

This variability of qubit noise profiles and the scaling challenges of deep learning models motivate the need for efficient alternatives. To develop such efficient noise protocols, we model a classical stochastic environment. In contrast, recent theoretical developments have characterised quantum environments without relying on classical stochastic realisations~\cite{winczewski2024intermediate, szankowski2017}. While classical noise produces mixed states via trajectory averaging, a quantum environment naturally captures more complex interactions such as non-unital energy relaxation through entanglement and back-action with the environment. However, a classical stochastic environment is the standard assumption in qubit control~\cite{ball2021software, cywinski2008enhance}, and keeps the model simple and computationally efficient. This simplicity aligns with the objective to provide a ``cheap and fast'' diagnostic tool suitable for practical, real-time noise identification on NISQ hardware, rather than an exhaustive quantum description of the environment.

We distinguish between three objectives: \textit{noise classification}, which assigns the noise to a broad physical category (e.g., $1/f$ versus coloured noise); \textit{noise identification}, which focuses on finding specific parameters (such as the location of a resonance peak) within an assumed model; and \textit{PSD reconstruction}, which involves mapping the continuous functional form of the noise spectrum. Traditional spectroscopy focusing on full PSD reconstruction drives up computational and experimental costs. Here, we shift the focus to classification and identification to enable efficient operations. 

For this classification and identification task, we introduce a quantum feature space ($\mathcal{F}$) that encodes the environment's influence on qubit dynamics, avoiding the overhead of continuous parameter extraction. We show that no expensive neural networks are needed; instead, simple machine learning algorithms, such as random forests, combined with a single pulse sequence, can quickly classify the noise affecting a qubit (\cref{fig:schematic}).

\begin{figure}[htbp]
    \centering
    \resizebox{\textwidth}{!}{
\begin{tikzpicture}[
 node distance=1.5cm and 1.0cm,
 box/.style={rectangle, draw=black!70, rounded corners, align=center, thick, minimum width=3.5cm, minimum height=1.2cm, fill=blue!5},
 arrow/.style={-{Stealth[scale=1.2]}, thick, draw=black!80},
 label/.style={align=center, font=\small, text width=2.5cm}
 ]
        
        \node[box, fill=orange!10] (sim) {Qubit Dynamics \\ $H(t) = H_{\mathrm{ctrl}}(t) + H_{\mathrm{noise}}(t)$};
        
        \node[box, fill=green!10, above left=1.0cm and 0.2cm of sim] (env) {Noise Environment \\ PSD $\to \beta_j(t)$};
        \node[box, fill=red!10, above right=1.0cm and 0.2cm of sim] (ctrl) {Control Pulses \\ $f_j(t)$ };
        
        \node[box, fill=purple!10, below=1.5cm of sim] (qfs) {Quantum Feature Space ($\mathcal{F}$) \\ $\mathbb{R}^9$ Representation};
        
        \node[box, fill=cyan!10, below=1.5cm of qfs] (ml) {Machine Learning \\ Decision Tree Partitioning};
        
        \node[box, fill=yellow!10, below=1.5cm of ml] (output) {Hardware Diagnosis \\ Noise Type, Stationarity, Freq.};

        \draw[arrow] (env) -| (sim);
        \draw[arrow] (ctrl) -| (sim);
        \draw[arrow] (sim) -- node[right, label] {Extract 9 Expectations} (qfs);
        \draw[arrow] (qfs) -- node[right, label] {Distance Mapping} (ml);
        \draw[arrow] (ml) -- (output);
        
        \draw[arrow, dashed, draw=gray, rounded corners=15pt] (output.east) -| node[right, label, pos=0.75] {Offline Control Optimisation} (ctrl.south);
        \draw[arrow, dashed, draw=gray, rounded corners=15pt] (output.west) -| node[left, label, pos=0.75] {Offline Noise Identification} (env.south);
        
    \end{tikzpicture}
}
    \caption{Schematic of the Quantum Feature Space ($\mathcal{F}$) diagnostic framework. Environmental noise realisations and targeted control pulses govern the qubit dynamics. The resulting expectation values for states prepared along orthogonal Bloch axes and Pauli observables are mapped into a structured 9-dimensional feature space ($\mathcal{F}$). This parameterisation allows efficient classical machine learning algorithms, such as Random Forests, to rapidly classify complex noise types and determine underlying physical parameters without the overhead of continuous spectrum reconstruction. This framework facilitates rapid offline noise identification, which can be leveraged for offline control pulse optimisation via digital simulations of a physical quantum computer.}
    \label{fig:schematic}
\end{figure}

\section{Deriving the Quantum Feature Space}

\label{sec:deriving-the-quantum-feature-space}

We model a single qubit perturbed by classical noise processes (using units where $\hbar=1$). The total Hamiltonian is given by
\begin{equation}
\label{eq:total_hamiltonian}
H(t) = H_{\mathrm{ctrl}}(t) + H_{\mathrm{noise}}(t),
\end{equation}
where the control Hamiltonian is
\begin{equation}
\label{eq:control_hamiltonian}
H_{\mathrm{ctrl}}(t)=\sum_{j \in \{x, y, z\}} f_{j}(t) \frac{\sigma_{j}}{2},
\end{equation}
and we operate in a rotating frame with $f_j(t)$ modelling externally applied control pulses. The qubit-environment interaction is defined as
\begin{equation}
\label{eq:qubit_environment_hamiltonian}
H_{\mathrm{noise}}(t) = \sum_{j \in \{x, y, z\}} \beta_{j}(t)\sigma_{j},
\end{equation}
where $\beta_j(t)$ is a stochastic process representing noise.

We are interested in the expected value of a qubit observable $O$ at the final time $T$ of the control pulse sequence, given an initial state $\rho$,
\begin{equation} \label{eq:initial_expectation}
\mathbb{E}\{O(T)\}_{\rho}=\left\langle\mathrm{Tr}\left[U(T)\rho U(T)^{\dagger} O\right]\right\rangle,
\end{equation}
where $U(T)=\mathcal{T} e^{-i \int_{0}^{T} H(s) \, ds}$ is the time-ordered evolution operator, and $\langle\cdot\rangle$ denotes classical averaging over the stochastic realisations of $\beta_{j}(t)$.

We initialise the qubit along the positive principal axes by writing $\rho_{j_+} = \frac{1}{2}(I + \sigma_j)$ with $j \in \{x, y, z\}$, and limit the observable to the Pauli operators, $\sigma_k$ with $k \in \{x, y, z\}$,
\begin{equation}
\mathbb{E}\{\sigma_k\}_{\rho_{j_+}} = \frac{1}{2} \left\langle \mathrm{Tr}\left[U (I + \sigma_j) U^{\dagger}\sigma_k\right] \right\rangle = \frac{1}{2} \langle \mathrm{Tr}[\sigma_k] \rangle + \frac{1}{2} \langle \mathrm{Tr}[U\sigma_j U^{\dagger}\sigma_k] \rangle,
\end{equation}
where, for brevity, we have omitted $T$, and operators are understood to be evaluated at the final time $T$ unless otherwise specified. Since $\mathrm{Tr}[\sigma_k] = 0$ for all $k \in \{x, y, z\}$,
\begin{equation}
\mathbb{E}\{\sigma_k\}_{\rho_{j_+}} = \frac{1}{2} \langle \mathrm{Tr}[U\sigma_j U^{\dagger}\sigma_k] \rangle,
\end{equation}
yielding nine independent measurements corresponding to $j, k \in \{x, y, z\}$. We then define the observable-specific feature vector as
\begin{equation}
\label{eq:quantum_feature_space_observable}
\mathbf{v}_{\sigma_k} = \left[ \mathbb{E}\{\sigma_k\}_{\rho_{x_+}}, \mathbb{E}\{\sigma_k\}_{\rho_{y_+}}, \mathbb{E}\{\sigma_k\}_{\rho_{z_+}} \right]^T. 
\end{equation}
The quantum feature space vector $\mathcal{F}$ is then defined as the concatenation of these vectors,
\begin{equation}
\mathcal{F} = \operatorname{vec}\left( [ \mathbf{v}_{\sigma_x}, \mathbf{v}_{\sigma_y}, \mathbf{v}_{\sigma_z} ] \right).
\end{equation}
This yields a 9-dimensional vector that captures the qubit-environment dynamics and incorporates hardware non-idealities, such as state preparation and measurement (SPAM) and control pulse errors, without the overhead of full state or process reconstruction. Importantly, rather than representing a static quantum state, $\mathcal{F}$ provides a compact, control-conditioned representation of the qubit-environment interaction.

We prepare only the positive eigenstates $\rho_{j_+} = \frac{1}{2}(I + \sigma_j)$ because the negative eigenstates $\rho_{j_-} = \frac{1}{2}(I - \sigma_j) = I - \rho_{j_+}$ yield redundant measurements under unital dynamics. Specifically, for traceless Pauli observables $O = \sigma_k$, any expectation measured from $\rho_{j_-}$ satisfies $\mathbb{E}\{\sigma_k\}_{\rho_{j_-}} = -\mathbb{E}\{\sigma_k\}_{\rho_{j_+}}$.

\subsection{Operator Interpretation \label{sec:operator_interpretation}}
Although $\mathcal{F}$ is defined in vector format for efficient storage and direct compatibility with classifiers, it admits a direct physical interpretation. For any unitary $U \in \mathrm{SU}(2)$, the trace inner product $R_{kj} = \frac{1}{2}\mathrm{Tr}[U\sigma_j U^{\dagger}\sigma_k]$ defines the $(k,j)$ entry of an $\mathrm{SO}(3)$ rotation matrix. Averaging over the stochastic realisations yields a $3 \times 3$ matrix with entries $T_{kj} = \mathbb{E}\{\sigma_k\}_{\rho_{j_+}}$, representing the ensemble-averaged rotation matrix of the channel. While each individual realisation is a unitary rotation, their ensemble average $T$ is a contractive matrix encoding decoherence. Consequently, the Euclidean distance between two feature vectors in $\mathcal{F}$ is identical to the Frobenius distance (and proportional to the Hilbert--Schmidt distance) between their corresponding $3 \times 3$ channel matrices.

\section{Numerical Simulation Implementation \label{sec:numerical_simulation_implementation}}
Our quantum simulator followed the formalism in~\cref{sec:deriving-the-quantum-feature-space} and was implemented natively in PyTorch~\cite{paszke2019pytorch}. The codebase is available at \url{https://github.com/ChrisWise07/quantum_feature_space}. Following~\cite{perrier2022qdataset}, we discretised the time interval $[0, T]$ into $M = 1024$ uniform timesteps of duration $\Delta t = T / M$, defining the discrete time grid $t_m = m \Delta t$ for $m \in \{0, 1, \dots, M-1\}$, with the total sequence duration normalised to $T = 1.0$.

\subsection{Creation of Noise Data \label{subsec:creation_of_noise_data}}
In our simulations, we implemented the following noise Hamiltonian~\cite{perrier2022qdataset},
\begin{equation}
\label{eq:noise_hamiltonian_in_sim}
H_{\mathrm{noise}}(t_m) = \beta_x(t_m) \sigma_x + |\beta_x(t_m)| \sigma_z,
\end{equation}
to couple both transverse and longitudinal noise to a common physical source.

Inspired by Perrier, Youssry, and Ferrie~\cite{perrier2022qdataset}, we explored three distinct physical noise profiles:
\begin{itemize}
\item \textbf{$\boldsymbol{1/f}$ noise:} Representing typical charge and flux noise fluctuations with power spectral density $\propto 1/f^{\alpha}$ commonly found in superconducting qubits~\cite{paladino20141}.
\item \textbf{$\boldsymbol{1/f}$ noise with a Gaussian bump:} Simulating a background $1/f^{\alpha}$ spectrum superimposed with a high-frequency resonance centred at frequency bin $\mu$, modelling a strongly coupled individual fluctuator~\cite{muller2019towards, krantz2019quantum}.
\item \textbf{Gaussian coloured noise:} Serving as a distinctly varied environment to test classification boundaries.
\end{itemize}

Gaussian coloured noise was generated by applying a moving-average filter of length $M / d$ to white Gaussian noise, where the division factor $d \in \{2, 3, 4, 5\}$ controls the correlation length. The Timmer--K\"onig method~\cite{timmer1995generating} was used for generating $1/f$ noise frequency-domain realisations, obtaining the time-domain sequences via inverse Fourier transform.

Similar to~\cite{perrier2022qdataset}, we created both stationary and non-stationary versions of each noise profile to test how non-stationarity manifests in $\mathcal{F}$ and affects classification performance. Non-stationarity was introduced by multiplying the stationary noise time series realisations by a Gaussian time envelope $g(t_m) = \exp[-(t_m - cT)^2 / (2\sigma_t^2)]$, where $\sigma_t = T/6$, with peak position $c$ sampled uniformly from $[0.2, 0.8]$.

Using this formulation, we generated 600 distinct noise processes (200 of each physical type, divided equally between stationary and non-stationary profiles). For each distinct noise process, an ensemble of up to $K=4096$ independent realisations was generated.

To guarantee a diverse and representative dataset, the defining parameters for each of the 600 noise profiles were randomised. \Cref{tab:noise_params} outlines the exact parameter bounds utilised in the dataset generation.

\begin{table}[htbp]
\centering
\begin{tabular}{lll}
\toprule
\textbf{Noise Profile Type} & \textbf{Parameter}           & \textbf{Range/Value} \\ \midrule

$1/f$                       & $\alpha$ (exponent)          & $[0.5, 1.5]$         \\

$1/f$ + Bump              & $\mu$ (bump frequency bin)  & $[5.0, 25.0]$        \\

& $\alpha$ (exponent)          & $[0.5, 1.5]$         \\

Coloured                    & $d$ (division factor)        & $\{2, 3, 4, 5\}$     \\

Non-Stationary              & $cT$ (envelope peak time)    & $[0.2T, 0.8T]$       \\ \bottomrule

\end{tabular}
\caption{Summary of bounds utilised to randomise the 600 distinct noise profiles. These bounds ensure spectral features lie within the dynamical decoupling filter passbands while providing sufficient variance.}
\label{tab:noise_params}
\end{table}

\subsection{Energy Normalisation} \label{sec:energy_normalisation}
To ensure that the classification models learnt underlying geometric and spectral structures encoded in $\mathcal{F}$, we prevented the models from trivially classifying datasets based only on total signal energy, defined as
\begin{equation}
\label{eq:signal_energy}
E_{\text{signal}} = \sum_{j \in \{x,y,z\}} \sum_{m=0}^{M-1} |\beta_j(t_m)|^2.
\end{equation}
To do this, we first determined the energy targets ($E_{\text{target}}$) for each of the six noise classes via a binary search protocol. That is, we iteratively scaled the noise profiles $\beta_j(t_m)$ until the uncontrolled FID of the qubit reached $1/e$ at the final sequence time $T$, at which point we recorded the corresponding total signal energy as $E_{\text{target}}$. Physically, this establishes that the sequence window corresponds to $T_2^*$ for all noise profiles, i.e., $T = T_2^*$.

Then, during data generation, we enforced these target energies across the entire ensemble by applying
\begin{equation}
\beta_j(t_m) \rightarrow \beta_j(t_m) \sqrt{\frac{E_{\text{target}}}{E_{\text{signal}}}}.  
\end{equation}
\subsection{Control Pulse Implementation}
Control pulses were applied solely along the $x$-axis ($f_y(t_m) = f_z(t_m) = 0$) for all simulations, taking the form of Gaussian-approximated CPMG sequences,
\begin{equation} \label{eq:gaussian_pulse}
f_{x}(t_m)=\sum_{p=1}^{N_p} (A_p \pm \nu_p) e^{-\frac{\left(t_m-\tau_{p}\right)^{2}}{2 \sigma^{2}}},
\end{equation}
where $\sigma$ is the Gaussian pulse width, $p \in \{1, \dots, N_p\}$ is the pulse index, and $\tau_{p}=\left(\frac{p-0.5}{N_p}\right) T \pm \delta_{p}$ sets the jittered pulse centre points. Here, $\delta_{p} \in \mathbb{R}$ explicitly models hardware jitter applied to the pulse timing. $A_p$ dictates the amplitude of the $p^{\mathrm{th}}$ pulse (which is dynamically scaled inversely to the pulse width $\sigma$ to maintain a constant pulse area of exactly $\pi$), and $\nu_p$ represents the dynamic amplitude jitter for the $p^{\mathrm{th}}$ pulse.

\begin{figure}[htbp]

\centering

\includegraphics[width=1.0\textwidth]{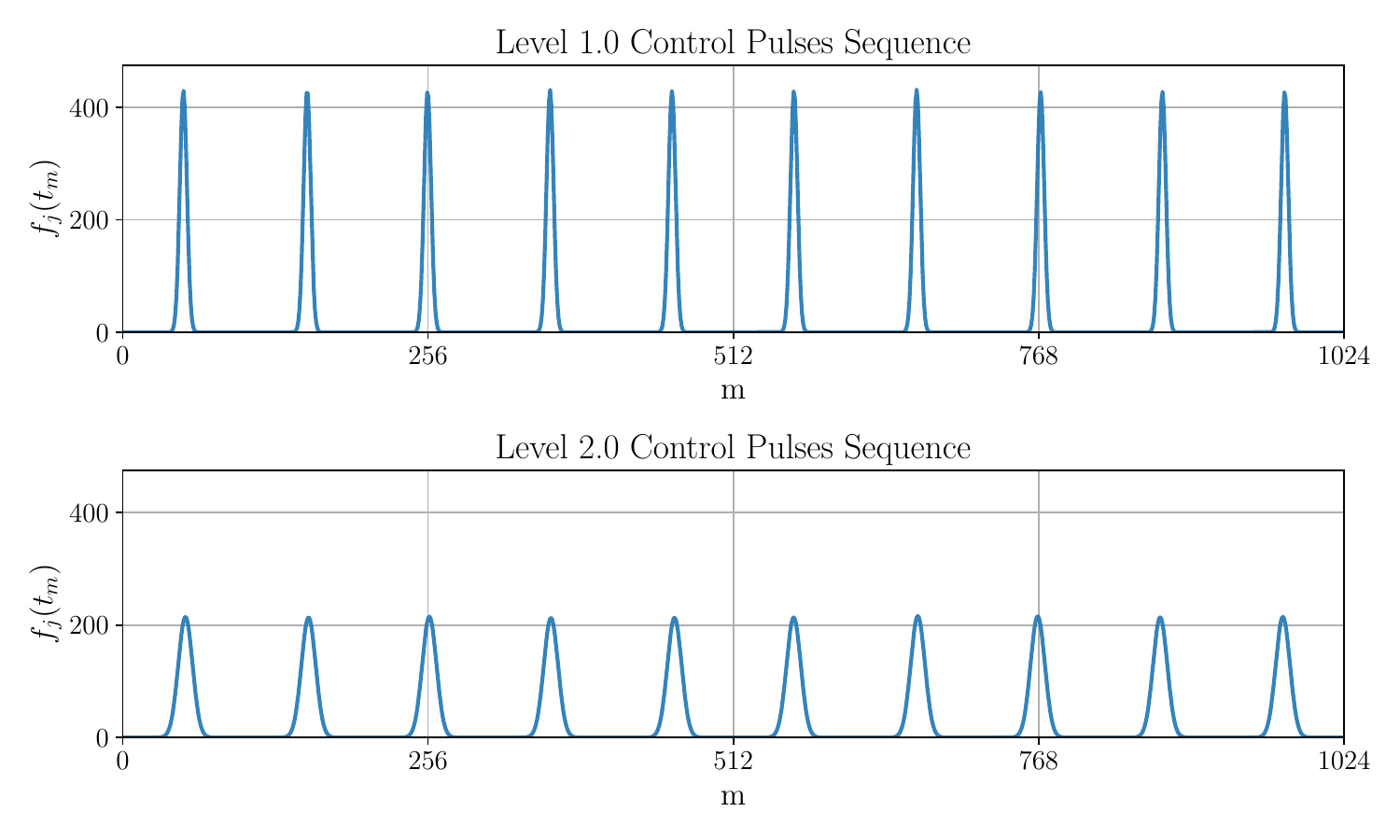}

\caption{CPMG-10 control pulse sequences generated for the simulations. (Top) Level 1.0 pulses with minor jitter. (Bottom) Level 2.0 pulses showcasing severe dynamic jitter in amplitude and timing across multiple realisations, representing highly degraded hardware imperfections. Note the reduction in peak amplitude for the wider level 2.0 pulses to preserve the $\pi$ pulse area.}

\label{fig:control_pulses}

\end{figure}

For our experiments, the sequence utilised $N_p=10$ $\pi$-pulses (a CPMG-10 sequence). For each control pulse level, we generated $K$ realisations by sampling the dynamic amplitude and timing jitter terms ($\nu_p$ and $\delta_{p}$) according to the hardware parameters in~\cref{subsec:hardware_assumptions}. \Cref{fig:control_pulses} illustrates level 1.0 and 2.0 pulse sequences with amplitude and timing jitter.

\subsection{Expectation Calculation}
We used~\cref{eq:gaussian_pulse} to generate $K$ control pulse sequence realisations $f_j(t_m)$ of length $M$, which were then used to construct the control Hamiltonians, $H_{\mathrm{ctrl}}(t_m)$, across all timesteps following~\cref{eq:control_hamiltonian}. We then selected $K$ noise realisations (of the 4096 total), each of length $M$, to construct the noise Hamiltonians, $H_{\mathrm{noise}}(t_m)$, across all timesteps following~\cref{eq:noise_hamiltonian_in_sim}. The total Hamiltonian $H(t_m)$ was then the sum of the control and noise Hamiltonians at each discrete timestep following~\cref{eq:total_hamiltonian}.

With the total Hamiltonian $H(t_m)$ defined at each discrete timestep, we computed the total evolution operator $U(T)$ using the Lie--Trotter product formula:
\begin{equation} \label{eq:lie-trotter-product-formula}
U(T) = \mathcal{T} e^{-i \int_0^T H(s) \, ds} \approx \mathcal{T}\prod_{m=0}^{M-1} e^{-i H(t_m) \Delta t} = e^{-i H(t_{M-1}) \Delta t} \cdots e^{-i H(t_0) \Delta t}.
\end{equation}

\Cref{eq:initial_expectation} was then used to compute the expectation estimates, and by exploiting the batched structure of the tensors, we could simultaneously compute all nine expectation values for a given $U(T)$. These expectation values were calculated by averaging over these $K$ classical noise and jitter realisations.

\section{Classification Results \label{sec:simulation_and_learning_results}}
\subsection{Hardware Assumptions and Error Modelling \label{subsec:hardware_assumptions}}
We set the baseline (level 1.0) parameters to represent standard superconducting transmon devices (e.g., IBM Quantum architectures)~\cite{krantz2019quantum, jurcevic2021demonstration}, with a baseline single $\pi$-pulse duration of $35\text{ ns}$. While physical transmons typically exhibit $T_2^*$ times of tens to hundreds of microseconds~\cite{krantz2019quantum, jurcevic2021demonstration}, we set $T = T_2^* = 2\ \mu\text{s}$. This choice is governed by numerical resolution: with $M = 1024$ timesteps, $T = 2\ \mu\text{s}$ provides a timestep of $\Delta t \approx 1.95\text{ ns}$, which is fine enough to resolve the narrowest $8.75\text{ ns}$ pulses (level 0.0) used in the experiments; in contrast, $T = 100\ \mu\text{s}$ would result in $\Delta t \approx 98\text{ ns}$, which is too coarse to resolve nanosecond pulse dynamics. Following the Nyquist--Shannon sampling theorem, these grid settings yield a Nyquist frequency of $f_{\text{max}} = \frac{M}{2T} = 256\text{ MHz}$ and a frequency resolution of $\Delta f = \frac{1}{T} = 0.5\text{ MHz}$.

\begin{table}[htbp]
\centering
\resizebox{\textwidth}{!}{%
\begin{tabular}{cccccccc}
\toprule
\multirow{2}{*}{\textbf{Level}} & \multicolumn{3}{c}{\textbf{Pulse Parameters}} & \multicolumn{2}{c}{\textbf{SPAM (\%)}} & \multirow{2}{*}{\textbf{Energy}} & \multirow{2}{*}{\textbf{Realisations ($K$)}} \\
\cmidrule(lr){2-4} \cmidrule(lr){5-6}
& \textbf{Width ($\sigma$)} & \textbf{Amp. Jitter ($\nu_p$)} & \textbf{Time Jitter ($\delta_p$)} & $\epsilon_0$ & $\epsilon_1$ & & \\
\midrule
0.0 & $8.75\text{ ns}$ & $0$ & $0$ & 0.0 & 0.0 & $0.25\times$ & 4096 \\
0.5 & $17.5\text{ ns}$ & $\pm0.0025\pi$ & $\pm1\text{ ns}$ & 1.0 & 1.5 & $0.50\times$ & 2048 \\
1.0 & $35.0\text{ ns}$ & $\pm0.0050\pi$ & $\pm2\text{ ns}$ & 2.0 & 3.0 & $1.00\times$ & 1024 \\
1.5 & $52.5\text{ ns}$ & $\pm0.0075\pi$ & $\pm3\text{ ns}$ & 4.0 & 6.0 & $1.50\times$ & 512 \\
2.0 & $70.0\text{ ns}$ & $\pm0.0100\pi$ & $\pm4\text{ ns}$ & 8.0 & 12.0 & $2.00\times$ & 256 \\
\bottomrule
\end{tabular}
}
\caption{Summary of experimental degradation levels simulating realistic hardware imperfections. The pulse amplitude jitter explicitly varies the target $\pi$ area, and the time jitter bounds are determined by the $200\text{ ns}$ pulse intervals. Level 1.0 represents standard NISQ baseline capabilities, while level 2.0 models severe calibration drift.}
\label{tab:hardware_levels}
\end{table}

\begin{itemize}

\item \textbf{Pulse Error Sensitivity:} We introduced Gaussian jitter to both the target pulse amplitude ($A_{p} \pm \nu_{p}$) and centre timings ($\tau_{p}=\left(\frac{p-0.5}{N_p}\right) T \pm \delta_{p}$), whilst widening Gaussian envelopes representing lower maximum amplitude allowed by the equipment. These errors were escalated across the 5 specific levels outlined in~\cref{tab:hardware_levels}.

\item \textbf{Noise Energy Sensitivity:} We applied a global scaling to the noise realisations via scalar multipliers, as outlined in~\cref{tab:hardware_levels}, to observe how varying the signal-to-noise ratio affects diagnostic performance.

\item \textbf{Realisation Sensitivity:} We varied the finite number of noise and control sequence realisations utilised for ensemble averaging. We mapped realisation numbers $K$ to the escalating levels in~\cref{tab:hardware_levels}.

\item \textbf{SPAM Errors:} We simulated SPAM errors via a transformation to the calculated expectations: $\tilde{\mathbb{E}}\{\sigma_k\}_{\rho_{j_+}} = \mathbb{E}\{\sigma_k\}_{\rho_{j_+}} \cdot (1 - \epsilon_0 - \epsilon_1) + (\epsilon_1 - \epsilon_0)$, where $\epsilon_0$ and $\epsilon_1$ denote the readout misclassification rates, with $\epsilon_0$ representing the probability of incorrectly reading a $\ket{0}$ as $\ket{1}$, and $\epsilon_1$ representing the probability of incorrectly reading a $\ket{1}$ as $\ket{0}$. The rates for $\epsilon_0$ and $\epsilon_1$ were scaled according to the levels detailed in~\cref{tab:hardware_levels}, with error rates modelling those discussed in~\cite{jurcevic2021demonstration, maciejewski2020mitigation}.

\item \textbf{Combined Errors:} We simultaneously tested all the above error sources by escalating them collectively according to the combined parameter sets defined in~\cref{tab:hardware_levels}. This evaluated the classifier's robustness under compounded experimental imperfections to indicate expected real-world performance.

\end{itemize}

\subsection{Robustness Tests}
Using the generated $\mathcal{F}$ parameter datasets, we evaluated a Random Forest classifier, K-Nearest Neighbours (KNN), and Logistic Regression algorithms. For all tests, the models were trained and evaluated using 10-fold cross-validation. The results displayed in the accompanying figures and text denote the mean accuracy across these folds, alongside the standard deviation.

For all three algorithms, we utilised the standard implementations provided by the Scikit-Learn library~\cite{scikit-learn}. The Random Forest classifier was constructed with default hyperparameters, comprising an ensemble of 100 decision trees governed by the Gini impurity split criterion. The K-Nearest Neighbours algorithm was similarly configured to its default of 5 nearest neighbours, evaluated using the standard Euclidean distance metric. The Logistic Regression model was configured with an L2 regularisation penalty ($C=1.0$) and allowed a maximum of 1000 iterations to ensure optimal convergence.

Across all tests, the Random Forest consistently achieved high test accuracy (averaging $\sim 89\text{--}90\%$) for classifying the noise profile type, demonstrating resilience to severe experimental constraints and imperfections. Furthermore, the Random Forest identified noise stationarity with near $100\%$ accuracy across all evaluations.

Logistic Regression underperformed compared to the Random Forest and KNN algorithms (e.g., dropping to $\sim 80\text{--}82\%$ for noise type, and exhibiting much higher variance). This provides empirical evidence that the mapping from the underlying noise parameters into the 9-dimensional $\mathcal{F}$ is non-linear, creating a data structure that the Random Forest partitions but linear decision boundaries cannot.

\subsubsection{Pulse Error Sensitivity}
We first tested the sensitivity of the classifiers to pulse errors. For these experiments, we used the following parameters:
\begin{itemize}
\item Noise Energy: $1.0\times$ (baseline)
\item Realisations: $K=4096$
\item SPAM Errors: level 0.0
\end{itemize}
where pulse errors varied from level 0.0 (no errors) to level 2.0 (severe errors) as detailed in~\cref{tab:hardware_levels}.

\begin{figure}[htbp]

\centering

\includegraphics[width=0.8\textwidth]{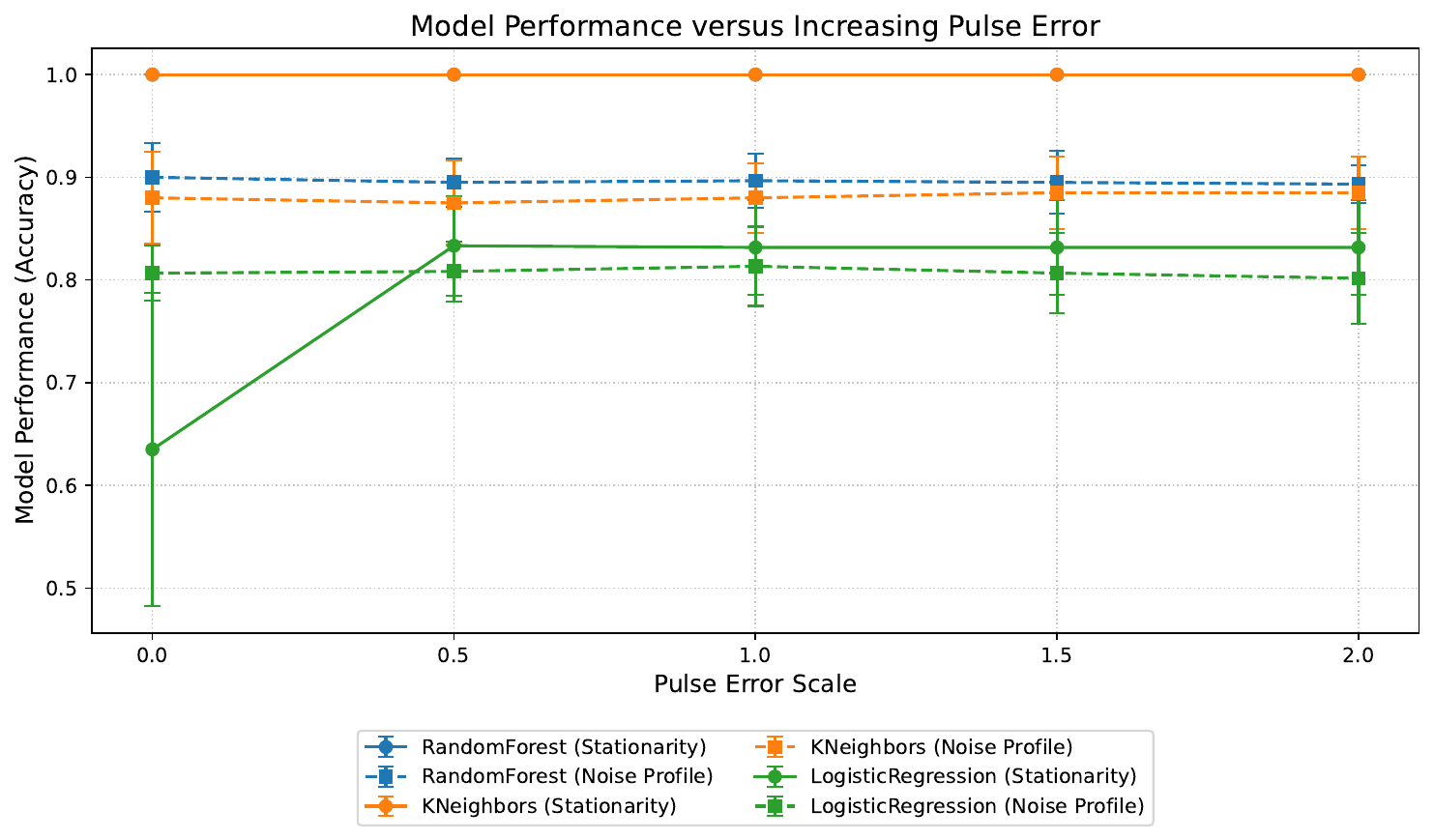}

\caption{Classifier performance under escalating levels of pulse amplitude jitter, timing errors, and Gaussian width expansion. The line displays the mean accuracy across the 10-fold cross-validation, and the shaded regions indicate the standard deviation.}

\label{fig:pulse_error}

\end{figure}

As shown in~\cref{fig:pulse_error}, varying the physical characteristics of the control pulses had minimal impact on the Random Forest and KNN classifiers. At the realistic standard IBM baseline (level 1.0), the Random Forest achieved a mean noise-type accuracy of $89.5\% \pm 2.4\%$. Even at the most severe level 2.0 degradation, classification accuracy was $90.2\% \pm 2.3\%$, reinforcing the robustness of the $\mathcal{F}$ representation against control pulse imperfections.

The Logistic Regression classifier for noise stationarity deteriorated at narrow pulse widths (particularly level 0.0), indicating possible sensitivity to narrow pulses. This trend continues across all other experiments involving narrow level 0.0 pulses.

\subsubsection{Noise Energy Sensitivity}
For the noise energy sensitivity tests, the following parameters were used:
\begin{itemize}
\item Pulse Errors: level 0.0
\item Realisations: $K=4096$
\item SPAM Errors: level 0.0
\end{itemize}
where the noise energy multiplier varied across the range $0.25\times$ to $2.0\times$.

\begin{figure}[htbp]

\centering

\includegraphics[width=0.8\textwidth]{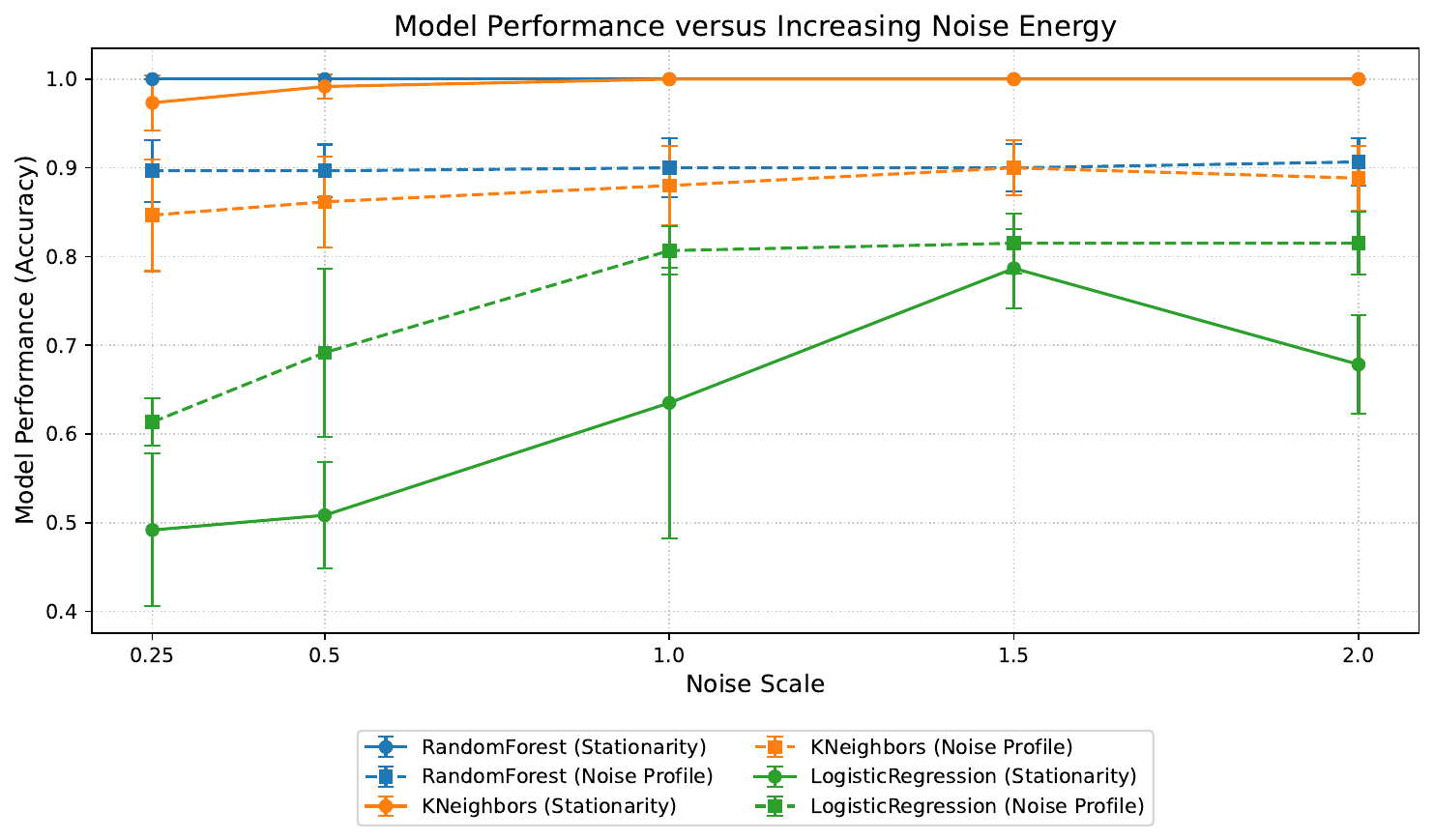}

\caption{Classifier performance across scaled environmental noise energy multipliers.}

\label{fig:noise_scaling}

\end{figure}

\Cref{fig:noise_scaling} demonstrates the impact of underlying noise power, and because the noise is pre-normalised to $T_2^*$, varying the energy tests the classifier's reliance on geometric structure rather than absolute amplitude. At the baseline $1.0\times$ multiplier, the Random Forest achieved $89.8\% \pm 3.2\%$ noise type accuracy. The Random Forest's accuracy remained high across both low-energy ($0.25\times$: $90.2\% \pm 3.0\%$) and high-energy ($2.0\times$: $90.8\% \pm 2.9\%$) regimes, demonstrating that the signature of the spectral features is retained within $\mathcal{F}$, independent of the signal magnitude.

KNN was moderately sensitive to noise energy variations, and Logistic Regression showed stronger sensitivity, with accuracy increasing as noise energy increased. This suggests that these algorithms may rely on the absolute amplitude of the noise, which could manifest in points separating further apart in the feature space as the noise energy grows.

\subsubsection{Finite-Ensemble Sensitivity}
\begin{figure}[htbp]

\centering

\includegraphics[width=0.8\textwidth]{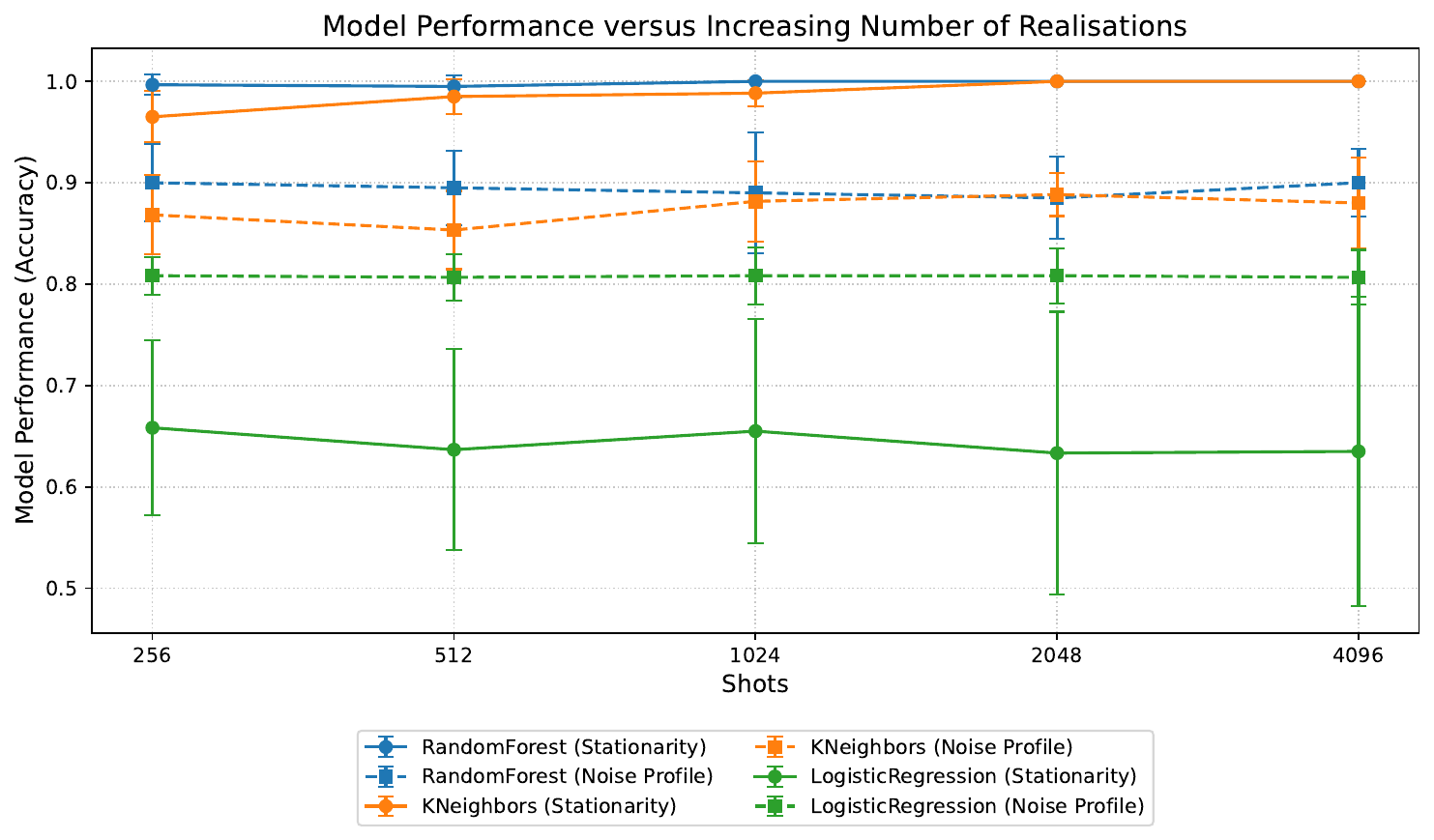}

\caption{Classifier accuracy plotted against the number of noise realisations ($K$) used for ensemble averaging to compute the observable expectations.}

\label{fig:shot_data}

\end{figure}

We then investigated the sensitivity of the classifiers to the number of stochastic noise realisations; to do so, we used the following parameters:
\begin{itemize}
\item Pulse Errors: level 0.0
\item Noise Energy: $1.0\times$ (baseline)
\item SPAM Errors: level 0.0
\end{itemize}
with the number of realisations varied as $K \in \{256, 512, 1024, 2048, 4096\}$. As illustrated in~\cref{fig:shot_data}, at $K=1024$ realisations, the model achieved a mean noise-type accuracy of $89.2\% \pm 5.5\%$. Decreasing the number of ensemble realisations from an ideal $K=4096$ down to $K=256$ realisations did not significantly degrade the Random Forest's predictive capability, going from $90.0\% \pm 3.0\%$ at $K=4096$ to $90.0\% \pm 4.0\%$ at $K=256$, meaning $\mathcal{F}$ extraction could be performed efficiently with limited measurement resources.

\subsubsection{SPAM Errors}
\begin{figure}[htbp]

\centering

\includegraphics[width=0.8\textwidth]{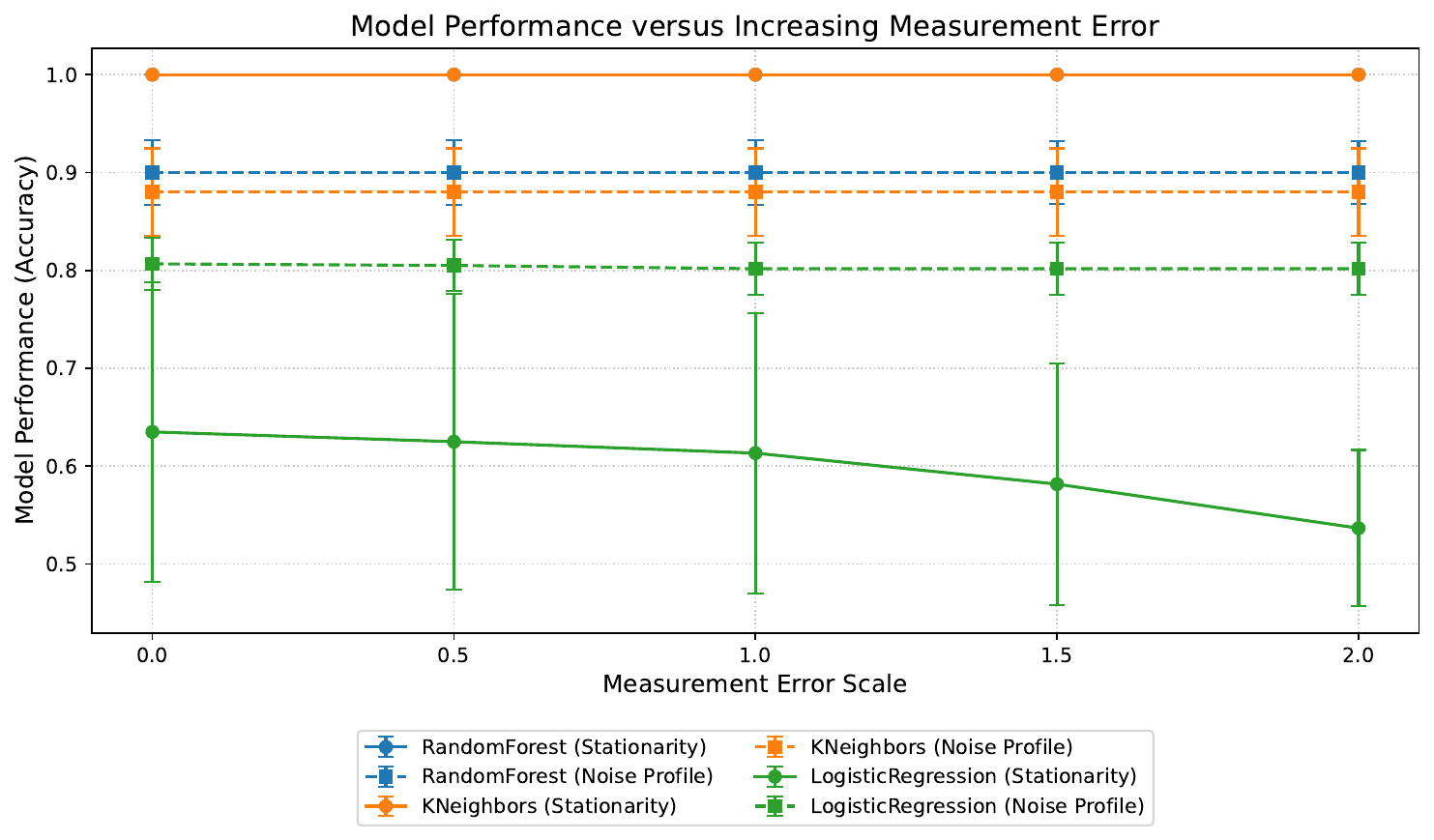}

\caption{Evaluation of the feature space under severe asymmetrical readout misclassification (SPAM) affine transformations.}

\label{fig:spam_error}

\end{figure}

To evaluate the impact of SPAM errors, we considered the following setup:

\begin{itemize}
\item Pulse Errors: level 0.0
\item Noise Energy: $1.0\times$ (baseline)
\item Realisations: $K=4096$
\end{itemize}
with SPAM errors varied from level 0.0 to level 2.0. As shown in~\cref{fig:spam_error}, all algorithms demonstrated resilience to SPAM errors (only the Logistic Regression algorithm diagnosing stationarity showed noticeable degradation). At the standard NISQ level 1.0 error rates ($\sim 2\text{--}3\%$ bit-flips), the Random Forest maintained $90.0\% \pm 3.3\%$ accuracy when diagnosing the noise profile. Despite introducing large readout bit-flips (averaging $10\%$ misclassification across the ground and excited states at level 2.0, with $\epsilon_0 = 8\%$ and $\epsilon_1 = 12\%$), the Random Forest was able to preserve high diagnostic fidelity ($90.0\% \pm 3.2\%$).

\subsubsection{Combined Errors}

Finally, to understand the likely performance of the QFS framework under realistic error conditions, we evaluated the models using the combined parameter sets defined in~\cref{tab:hardware_levels}. In the preceding tests, we isolated individual error sources by holding all other parameters at their baseline values (e.g., $K=4096$, $1.0\times$ noise scaling, and zero pulse/SPAM errors for level 0.0). However, in this combined test, we simultaneously escalated all sources of degradation. This meant moving from the most pristine quantum computer at level 0.0 (idealised $K = 4096$ realisations, no pulse errors, no SPAM errors, with a $0.25\times$ noise scaling) down to a severely degraded level 2.0 scenario (encompassing wide pulses, significant jitter, a reduced number of realisations $K=256$, a $2.0\times$ noise scaling, and $10\%$ SPAM bit-flips).

\begin{figure}[htbp]

\centering

\includegraphics[width=0.8\textwidth]{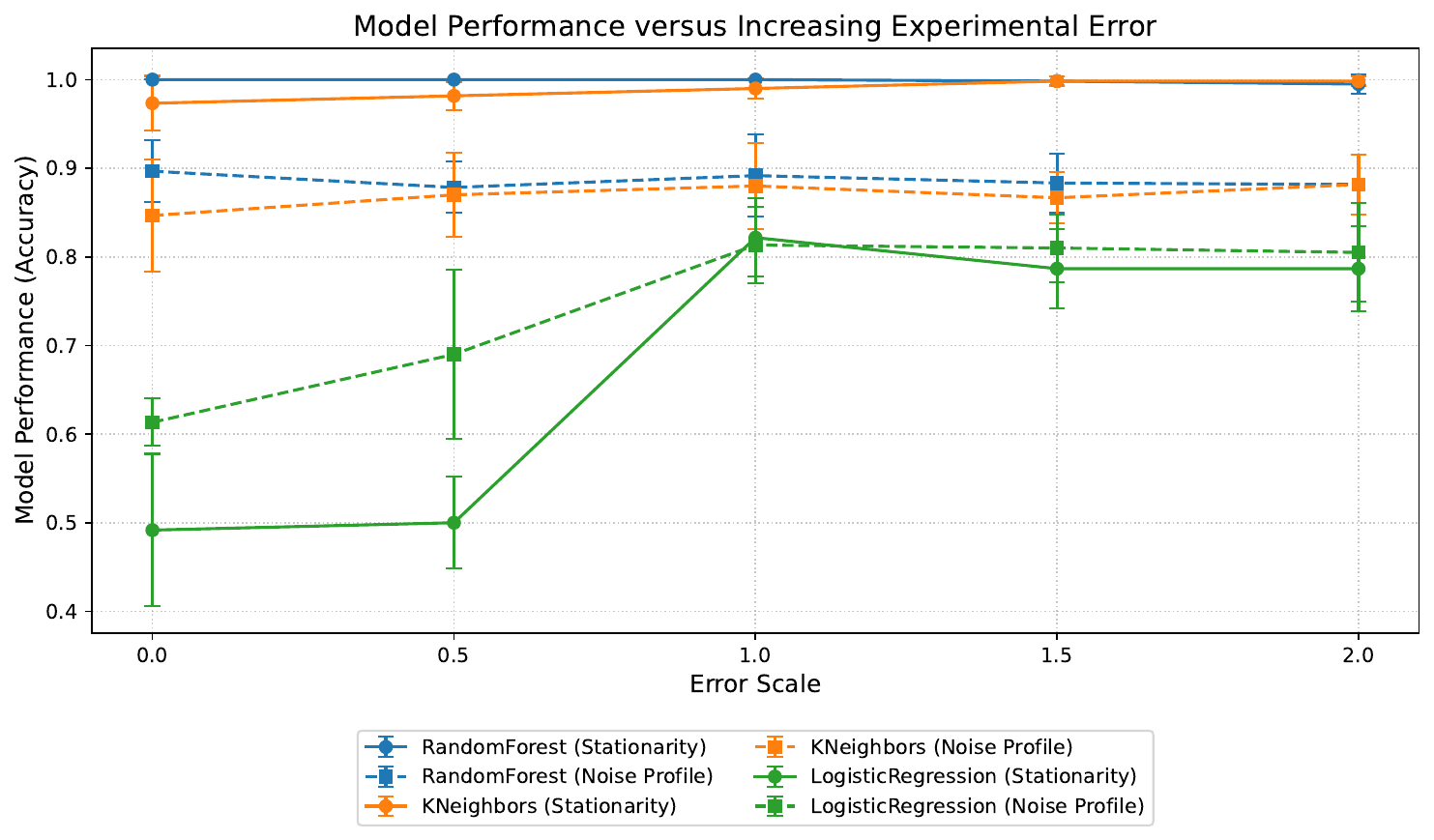}

\caption{Classifier performance when subjected to all experimental degradations and hardware constraints simultaneously.}

\label{fig:combined_error}

\end{figure}

\Cref{fig:combined_error} demonstrates the practical utility of the QFS framework. Despite increasing error, the Random Forest continued to correctly diagnose the underlying noise family. At the realistic level 1.0, the model achieved $89.2\% \pm 4.6\%$ for noise profile diagnosis, while maintaining a perfect $100.0\% \pm 0.0\%$ stationarity diagnosis. Even under the most severe error conditions at level 2.0, the Random Forest still achieved a diagnostic accuracy of $88.2\% \pm 3.3\%$.

The KNN and Logistic Regression also demonstrate a level of resilience, albeit with lower overall accuracy compared to the Random Forest, even under the combined severe error conditions.

\subsubsection{Summary of Robustness Tests}
Across all isolated and combined hardware degradations, the 9-dimensional $\mathcal{F}$ combined with ML classifiers, particularly the Random Forest, demonstrated robustness and reliability in diagnosing the underlying noise characteristics despite severe NISQ hardware limitations. The Random Forest model consistently delivered $\sim 90\%$ classification accuracy for the underlying physical noise type and $\sim 100\%$ accuracy for stationarity. The fact that the Random Forest successfully parsed $\mathcal{F}$ under combined, severe NISQ conditions (level 2.0) demonstrates that a single-pass extraction of the expectation values contains sufficiently robust structural information for practical downstream diagnostic tasks.

Furthermore, we speculate that all algorithms performed better for stationarity classification because of the energy normalisation. That is, because a non-stationary signal is characterised by a Gaussian bump where it is ``active'', it inherently possesses regions of near-zero amplitude at the edges of the sequence window. To achieve the exact same total sequence energy as a continuously active stationary signal, the non-stationary signal must exhibit a significantly higher peak variance during its active window. This pronounced difference in variance likely translates directly into distinct geometrical divergence within $\mathcal{F}$. Consequently, separating stationary from non-stationary processes becomes an inherently easier classification task.

\subsection{Black Box Classification of Noise Environment Through QFS \label{subsec:black_box_noise_spectroscopy}}
To demonstrate $\mathcal{F}$ as a practical and efficient diagnostic tool, we conducted a proof-of-concept black box classification experiment. We utilised the Random Forest trained explicitly on the level 1.0 Combined Error dataset, as this level provided the most realistic representation of standard NISQ hardware capabilities.

We started by modelling an ``unknown'' physical process as a $1/f$ noise spectrum with a resonance bump at a frequency bin of $\mu = 15.0$. The data point in $\mathcal{F}$ corresponding to this unknown environment was successfully diagnosed by the Random Forest as a stationary `$1/f$ + Bump' noise type.

Having identified the broad class of noise, we then aimed to identify the precise resonance frequency. Instead of executing a complex analytical spectral reconstruction, we generated a library of candidate `$1/f$ + Bump' noise profiles spanning various frequency bumps ($\mu \in [5.0, 25.0]$). To isolate the geometric behaviour of $\mathcal{F}$ to the spectral parameter $\mu$, the candidates were simulated using common random numbers where appropriate (i.e., the same control pulse realisations and noise seeds). Consequently, at the true candidate $\mu = 15.0$, the distance is $0.0$. 

\begin{table}[htbp]
\centering
\begin{tabular}{cccc}
\toprule
\textbf{Candidate Freq.} ($\mu$) & \textbf{Distance in $\mathcal{F}$} & \textbf{Candidate Freq.} ($\mu$) & \textbf{Distance in $\mathcal{F}$} \\
\midrule
5.0 & 0.032155 & 16.0 & 0.000251 \\

6.0 & 0.023452 & 17.0 & 0.000451 \\

7.0 & 0.016232 & 18.0 & 0.000624 \\

8.0 & 0.010713 & 19.0 & 0.000778 \\

9.0 & 0.006785 & 20.0 & 0.000906 \\

10.0 & 0.004152 & 21.0 & 0.001014 \\

11.0 & 0.002467 & 22.0 & 0.001085 \\

12.0 & 0.001411 & 23.0 & 0.001120 \\

13.0 & 0.000748 & 24.0 & 0.001132 \\

14.0 & 0.000312 & 25.0 & 0.001124 \\
\textbf{15.0} & \textbf{0.000000} & & \\
\bottomrule
\end{tabular}
\caption{Euclidean distance in the Quantum Feature Space between the unknown target ($\mu=15.0$) and various candidate frequency bumps.}
\label{tab:qfs_distances}
\end{table}
As recorded in~\cref{tab:qfs_distances}, the Euclidean distance is smallest at the correct candidate $\mu = 15.0$ (distance $\approx 0.0$), correctly identifying the spectral parameters of this unknown noise profile.

The distances exhibit a smooth, nearly monotonic relationship: as the candidate frequency moves away from the target $\mu=15.0$ in either direction, the Euclidean distance in $\mathcal{F}$ increases across almost the entire candidate range. This monotonic gradient structure suggests that the QFS maps physical parameters (like frequency) smoothly into distance, making this framework highly suitable for optimisation algorithms (like gradient descent or binary search) to rapidly fit noise parameters without exhaustive sweeps.

Here we employ the Euclidean distance over $\mathcal{F}$ as a simple, efficient metric that, as established in~\cref{sec:operator_interpretation}, is proportional to the Hilbert--Schmidt distance (and identical to the Frobenius distance) between the effective channel matrices. Investigating whether non-linear operator metrics, such as trace distance, Bures distance, or process fidelity, could further improve discrimination between candidate noise profiles remains an interesting direction for future work.

\section{Comparison to Previous Work \label{subsec:explicit_comparison_to_previous_work}}
The primary innovation of the QFS method is its substantial reduction in both model complexity and data requirements, directly addressing the key scaling limitations of deep learning strategies~\cite{wise2021using, martina2023deep, youssry2020characterization}. As shown in~\cref{tab:method_comparison_final}, deep learning models employ neural networks with high parameter counts (often $>10^4$ to $10^5$) that require thousands of data points for training. By contrast, the Random Forest used in our experiments has 8,192 parameters and used only 600 training examples.

Previous black-box and grey-box methods utilised neural networks to directly output full continuous PSD reconstructions or noise operator estimates~\cite{wise2021using, youssry2020characterization}. Because full continuous spectral mapping is an inherently difficult, high-dimensional task, these models are computationally expensive and require large training datasets. Instead, leveraging $\mathcal{F}$---which intrinsically encodes numerous dynamical properties of the qubit-environment interaction---allows us to break the characterisation process into two simplified phases. First, by applying simple ML algorithms to $\mathcal{F}$, we solve the easier problem of noise classification. Once the overarching noise class is known, we can use simulated data to test various parameters within that specific class in parallel, finding the parameter set that best matches the experimental $\mathcal{F}$ data. This two-phase approach allows for an accurate, targeted diagnosis of the specific noise spectra while reducing the computational and experimental cost. 

This work has not exhaustively stress-tested this parameter-fitting approach for full continuous PSD curve reconstruction across an arbitrary variety of complex spectra. However, the results here show that this efficient protocol of classification followed by targeted identification is a promising approach for downstream quantum optimal control, which is orders of magnitude faster than direct neural network approaches.

\begin{table}[htbp]
\centering
\resizebox{\textwidth}{!}{%
\begin{tabular}{l >{\raggedright\arraybackslash}p{3.0cm} >{\raggedright\arraybackslash}p{3.0cm} >{\raggedright\arraybackslash}p{3.0cm} >{\raggedright\arraybackslash}p{3.0cm}}
\toprule
\textbf{Attribute} & \textbf{FTNS \cite{vezvaee2024fourier}} & \textbf{NN Regression \cite{wise2021using, martina2023deep}} & \textbf{Grey-box ML \cite{youssry2020characterization}} & \textbf{QFS (This Work)} \\
\midrule
\textbf{Principle} & Analytical inversion & Deep NN & Deep NN + analytical layers & Euclidean distance / Random Forest \\
\textbf{Exp. Overhead} & FID curve (time sweep) & Decay curves ($\tau$ sweeps) & Single sequence & Single fixed sequence \\
\textbf{Comput. Cost} & Analytical (very low) & NN training (high) & NN training (high) & Distance calculation / Tree training (low) \\
\textbf{Fidelity/Accuracy} & $\sim 99\%$ & $\sim 95\text{--}98\%$ & $> 99\%$ & $\sim 90\%$ / Exact* \\
\textbf{Model Parameters} & N/A & $\sim 10^4\text{--}10^5$ & $\gtrsim 10^5$ & $\sim 8 \times 10^3$ \\
\bottomrule
\end{tabular}
}%
\caption{Comparison of qubit noise spectrum estimation methods. While previous methods focus on continuous PSD reconstruction, the QFS approach is highlighted for its ability to classify noise types and identify spectral features with low model complexity, making it highly suitable for rapid downstream diagnostics. *$\sim 90\%$ broad classification accuracy under level 1.0 combined errors; resonance peak $\mu$ identified exactly on the candidate grid.}
\label{tab:method_comparison_final}
\end{table}

\section{Scaling \label{sec:scaling}}
To scale the feature space to a two-qubit gate protocol, the qubit state and observables are represented by $4 \times 4$ matrices that span a 16-dimensional operator space (the generalised Pauli basis). Following the same logic as the single-qubit derivation, removing the uninformative identity terms requires $15 \times 15 = 225$ independent expectation measurements. While each configuration requires only a single-pass pulse sequence (not the exhaustive frequency sweeps of traditional spectroscopy), one would then need to classify the behaviour of all physical two-qubit gates across the hardware.

Because physical quantum computers decompose all operations into fundamental single- and two-qubit physical gates, the noise characterisation problem inherently simplifies into local components, avoiding the need to consider the full exponential scaling of the global Hilbert space.

For single-qubit gates, experimental data could be gathered for all physical qubits simultaneously, with the corresponding classical diagnosis computed in parallel using GPU acceleration.

However, diagnosing every physical pair of qubits introduces an experimental overhead as the total number of connected pairs grows rapidly, particularly in globally connected hardware topologies. This challenge can be mitigated with data collection parallelised across spatially disjoint qubit pairs. 

In graph-theoretic terms, characterising the noise of a fully connected $N$-qubit processor is equivalent to finding the 1-factorisation (or minimum edge colouring) of a complete graph $K_N$~\cite{west2001introduction}. For an even number of qubits, the $N(N-1)/2$ total two-qubit connections can be partitioned into exactly $N-1$ disjoint matchings. This guarantees that even in the worst-case scenario of a globally connected macroscopic processor, the required number of sequential measurement rounds scales strictly linearly. 

For example, fully characterising a 10-qubit processor (45 total pairs) requires 9 parallelised rounds, and a 100-qubit processor (4,950 total pairs) requires 99 rounds. Therefore, fully mapping two-qubit gate noise across a macroscopic NISQ processor avoids exponential bottlenecks, requiring at most $N$ parallelised rounds of experimental data collection (and exactly $N - 1$ when $N$ is even).

\section{Conclusion}
In this work, we introduced a quantum feature space ($\mathcal{F}$) constructed from a single pulse sequence, designed to intrinsically encode the qubit-environment interaction dynamics of a qubit subjected to a classical stochastic environment. By shifting the objective from computationally expensive, continuous full-spectrum reconstruction to targeted discrete classification, we demonstrated that the $\mathcal{F}$ framework eliminates the need for both exhaustive multi-pulse frequency sweeps on quantum hardware and expensive neural networks for classical processing.

Instead, we showed that applying simple, efficient ML algorithms, such as a Random Forest containing fewer than $10^4$ parameters, to $\mathcal{F}$ provides a robust and scalable diagnostic tool. Under simulated realistic hardware constraints incorporating pulse jitter and readout errors, our method successfully classified broad noise processes with $\sim 90\%$ accuracy. Furthermore, using simple Euclidean distance metrics over $\mathcal{F}$, we correctly identified spectral resonances. This protocol provides a rapid, computationally cheap diagnostic tool for real-time noise identification on macroscopic NISQ hardware.

Looking forward, this paper opens several promising avenues for future work. First, extending the framework to study non-Gaussian environments, such as random telegraph noise, will address strongly coupled two-level fluctuators common in solid-state devices.

Additionally, extending the framework to capture non-unital dissipative dynamics---where the Bloch sphere undergoes an affine shift and longitudinal relaxation towards a ground or thermal state rather than purely unital dephasing---is an important open direction. This could be achieved by coupling the qubit to a truncated quantum environment, which naturally produces non-unital relaxation through entanglement and energy exchange with the environment. Alternatively, for simpler diagnostic goals, the classical stochastic model could be phenomenologically augmented with an explicit amplitude damping channel to model $T_1$ relaxation. Any of these extensions would broaden the diagnostic scope of $\mathcal{F}$ to encompass the full range of decoherence mechanisms encountered on real hardware. 

Future work could examine the performance of ML classifiers as the number of timesteps $M$ changes, for example $M \in \{256, 512, 1024, 2048, 4096\}$. For a fixed sequence duration $T$, this evaluates classifier robustness across different noise spectral bandwidths and control pulse discretisation fidelities.

There is also significant potential for direct parameter estimation. By utilising PyTorch's native autograd capabilities, researchers could reparameterise the problem and backpropagate directly through the differentiable $\mathcal{F}$ mapping to estimate experimental noise parameters without relying on heuristic searches. This could involve parameterising the PSD as a weighted sum of basis functions, with the weights treated as optimisable parameters. This approach fuses noise classification and identification, potentially avoiding the need for any ML algorithms.

Applying the framework directly to physical NISQ hardware will be a vital next step. Executing these sequences on physical quantum processors will benchmark real-world fidelities and validate the resilience of the QFS against unmodelled environmental crosstalk.

To further enhance the practical utility of the QFS, future work could also study the robustness of the framework when extending to two qubits, where correlated noise and entanglement effects become significant.

Finally, utilising the established QFS as a lightweight digital twin presents an opportunity for advanced control. By mapping hardware responses to $\mathcal{F}$, we could backpropagate and optimise quantum control pulse sequences natively on classical GPUs. This would entirely bypass executing online hardware optimisations, enabling rapid, closed-loop pulse generation adapted to drifting environmental noise.

\section{Data and Code Availability \label{sec:data_and_code_availability}}
The data and code are available at \url{https://github.com/ChrisWise07/quantum_feature_space}.

\section{Funding \label{sec:acknowledgments}}
MW acknowledges support from the Australian Research Council (ARC) Centre of Excellence for Engineered Quantum Systems (CE170100009). AP acknowledges an RMIT University Vice-Chancellor's Senior Research Fellowship and a Google Faculty Research Award. This work was supported by the Australian Government through the Australian Research Council under the Centre of Excellence scheme (CE170100012).

\bibliography{references}

\end{document}